\documentstyle[aps,preprint,12pt]{revtex}
\begin{document}
\draft
\title{
Enhancement of Persistent Currents by Hubbard Interactions In
Disordered 1D Rings: Avoided Level Crossings Interpretation} 

\author{Michael  Kamal}

\address{
Goldman, Sachs \& Co., 85 Broad St.,
New York, NY 10004
}
\author{Ziad H. Musslimani and Assa Auerbach}
\address{
Department of Physics, Technion-Israel Institute of Technology\\ Haifa 32000,
 Israel}

\date{\today}
\maketitle
\begin{abstract}

We study effects of local electron interactions on the persistent current of
one dimensional disordered rings. For different realizations of disorder we compute the current as a function of Aharonov-Bohm flux to zeroth and first orders in the Hubbard interaction.
We find  that 
the persistent current is {\em enhanced} by onsite interactions.
Using an avoided level crossings approach, we derive analytic formulas
which explain the numerical results at weak disorder. The same approach also explains the opposite effect (suppression) found for spinless fermion models with intersite interactions.
\end{abstract}

\section{Introduction}
\label{sec:level1}

Recent experiments  have found that small isolated
metallic rings, threaded by magnetic flux, carry persistent currents.  
Although the existence of this effect had been anticipated theoretically
for many years 
\cite{yang,bloch,butt,cheung88}, the magnitude of the observed current was
found to be much larger than expected.  

Experiments performed on moderately disordered rings with small transverse
width have measured the magnetic response of an ensemble of copper rings
\cite{bell},
of isolated gold loops \cite{ibm}, and of 
a clean, almost
one-dimensional, sample  \cite{frenchguys}.

The ensemble averaged persistent
current was found to have an amplitude of order $ 10^{-2} ev_f/L $
and periodicity $ \Phi_0/2 $, (where $L$ is the circumference of the ring, $v_f$ is the Fermi velocity, $e$ is the electron charge, and $\Phi_0=h/e$ is the flux quantum).
The single loops experiment reported finding
an unexpectedly large current   of order $ev_f/L $, the free electron value,
and   displaying periodicity in flux with period $\Phi_0$.

Theoretical approaches can be classified into non-interacting and interacting electrons approaches. Non interacting theories which take into account
the effects of disorder and deviations from a perfect 1D geometry\cite{average}, predict 
values of the average persistent current  that are  smaller than observed, but 
still within one or two  orders of magnitude of the experimentally determined 
value
of the average current
. More surprisingly,  the theoretical 
estimates for the typical value 
of the persistent current, measured in  the single loop experiment, are perhaps
two or  three orders of magnitude smaller than the reported values. 
It seems that agreement between theory and experiment worsens with increasing
disorder. This suggests that  perhaps interaction corrections are quantitatively important
in the moderately disordered regime.

A first guess would be that interactions enhance the impurity scattering
and  further {\em suppress} the persistent current, as happens for the conductance of the disordered luttinger liquid \cite{Kane}. However, this intuitive analogy  may be quite misleading.  While the conductance depends 
on the Fermi surface properties, i.e. velocity and scattering rates, 
the persistent current is a sum of
contributions from all occupied states. Also, by Galilean invariance, an interacting electron liquid without disorder carries
the same current as the non interacting gas, even though its Fermi velocity
may be  renormalized.
Thus it is plausible that in the disordered system, interactions could work both ways, i.e. either suppress or enhance the persistent currents under conditions which need to be explored.

Exact diagonalization studies of spinless fermions with intersite interactions on  small rings \cite{berkovits,nnexact}, have found that weak 
interactions
{\em reduce} the persistent current below its  noninteracting value.
On the other hand, recent reports \cite{shastry,RRB} for  Hubbard interactions found the opposite effect. 

In this paper \cite{Kamal} we clarify this seeming contradiction, and gain some insight
into the puzzle of interaction corrections. In Section \ref{sec:numerical} we diagonalize the tight binding Hamiltonian numerically for different realizations of disorder, and compute the  persistent current to zeroth and to first order in the  Hubbard interactions.
We find  that interactions  {\em enhance} the persistent current. In Section \ref{SecALC} we present the  Avoided Levels Crossing (ALC) theory, which provides an analytic approximation to the numerical results. This is 
an expansion, at weak disorder, of nearly degenerate eigenstates at fluxes
close to points of time reversal symmetry. The correlation between the zeroth and first order currents is explained by a common mechanism, i.e. the avoided
level crossings. The opposite (suppression) effect for interacting
spinless fermions, is also explained by the
ALC theory in Section \ref{Sec4}. We conclude by a summary and future directions.

\section{Numerical Perturbation Theory}
\label{sec:numerical}

We consider a tight-binding Hamiltonian on a periodic chain with repulsive
on-site Hubbard interaction: 
\begin{equation}
 H= H_{0} \; + \;  U \sum_{i} n_{i\uparrow} n_{i\downarrow},
\end{equation}
\begin{equation}
\label{hamiltonian}
 H_{0}= - \sum_{i \sigma}[  e^{i( \frac{ 2 \pi} {L} \frac{\Phi} {\Phi_0} )} 
a_{i+1\sigma}^{\dagger}  a_{i\sigma}+{\rm h.c.}]
\; + \; \sum_{i \sigma} \epsilon_{i} a_{i\sigma}^{\dagger} a_{i\sigma}. 
\label{h0}
\end{equation}
where $a^\dagger_{i\sigma}$ creates an electron at site $i$ with spin $\sigma$.
Our unit of energy is the hopping energy. $\epsilon_i$ is the dimensionless on-site disorder energy
uniformly distributed in the interval [$-W/2$,$W/2$].
The chain has $L$ sites, $N_e$ electrons, the flux through its center 
is given by $\Phi$. 
With these conventions, the energy spectrum is
periodic in the enclosed flux with period $\Phi_{0}$,   and the current is given by
\begin{equation}
I(\Phi)= - \frac{\partial}{\partial \Phi}E(\Phi),
\end{equation}
where
\begin{equation}
E(\Phi)=  E_0(\Phi) +  E_1(\Phi)+ {\cal O}(U^2).
\end{equation}
$E_0$ is the exact ground state energy of $H_0$, whose single electron
eigenstates are determined numerically.  $E_1$ is the first order correction
in $U$, which is given by
\begin{eqnarray}
\label{e1def}
E_1&=& U\sum_{i}  n_{i\uparrow} n_{i\downarrow} \nonumber\\
n_{i\sigma}&=&\langle \Psi_0|a^\dagger_{i\sigma}a_{i\sigma}|\Psi_0\rangle
\end{eqnarray}
where $\Psi_0$ is the Fock ground state of $H_0$.
Thus by diagonalizing $H_0$ we can readily obtain
\begin{equation}
I(\Phi)=I_0(\Phi)+I_1(\Phi)+ {\cal O}(U^2).
\end{equation}

The effects of disorder on  $I_0(\Phi)$ can be seen in Fig.~1, where
it is shown as a function of flux for a half filled lattice of 
six sites, for different disorder strengths $W$. Disorder smoothens and reduces the magnitude of $I_0$, and for $W>>1$ 
it is dominated by the first harmonic $\sin(2\pi\Phi/\Phi_0)$. 

In Fig.~2, 
$I_{1}(\Phi)$ is plotted.
There are several features of the first order interaction correction 
that deserve comment.
First, the most important observation is that it 
{\it generally}  enhances the noninteracting current for all 
values of disorder strength and flux. 
That is to say: there is a {\em positive}
correlation between the non interacting current and the first order correction,
\begin{equation}
\langle I_0(\Phi) I_1(\Phi)\rangle_{\epsilon,\Phi} ~\ge 0.
\label{poscor}
\end{equation}
This result is found numerically to 
hold for all realizations of disorder which we have used. 
Second, in the limit of weak disorder,
$I_1(\Phi)$ becomes singular at discontinuity points  of
$I_0(\Phi)$, which are at fluxes $(m+{1\over 2} )\Phi_0$, ($ m \Phi_0$)
for even (odd) number of filled orbitals, where $m$ are integers.
Finally, we see the first order current also becomes dominated by its first harmonic 
at large disorder.

To study the scaling properties of the current with  system size
and disorder strength, it is convenient to characterize the strength of
both $I_0$ and $I_1$ by their amplitude at $\Phi=\Phi_{0}/4$.  
And while it
does not capture  the singular  nature of $I_{1}(\Phi)$ near the regime
of very weak disorder, this characterization is still useful  to study the
scaling properties of the first order current, even in this limit.  
Previous studies have found \cite{cheung88,berkovits} both numerically and analytically, that 
the amplitude
of the noninteracting current, averaged over disorder, behaves like:
\begin{equation}
\label{i0amp}
 |I_{0} (\Phi_{0}/4)| \simeq \frac {1} {2} \frac{ ev_{f} } {L}
\exp (-L/\xi). 
\end{equation}
By fitting 
to this functional form
we extracted the behavior of the localization length $\xi$ for
different system sizes and disorder values.  Figure \ref{loclength} shows
the values of the localization length  obtained for sizes $L$=6,10,14,20 and
25, at half-filling, averaged over many  (up to four thousand) 
impurity configurations for each value of $W$.  We find good agreement between 
our inferred value of the 
localization  length and the known asymptotic form  in the  weak disorder 
limit: 
$\xi=105/W^{2}$,  valid  when  $W\ll 2\pi$ for large systems at half-filling.

There is less agreement in the strongly localized limit where 
the localization length is known to  behave like: $\xi= (\ln(W/2)-1)^{-1}$.  
We find a better
fit taking $\xi=1/\ln(W^{0.6}/2.5)$  for the $L=25$ data.  This discrepancy
could be due to the small sizes considered, or a breakdown of  
Eq. (\ref{i0amp}) when $\xi$ is of order unity. 
We have calculated 
the amplitude $I_{1}(\Phi=\Phi_{0}/4)$  as function of strength
of disorder, characterized by the scaling parameter  $L/\xi$, for 
system sizes of $L$=6,10,14,20,25.
At  weak disorder (large $\xi$), the amplitude
increases with the strength of disorder achieving its maximum 
value at some intermediate strength of disorder ( $ \xi \simeq L$ ).  In this 
weakly disordered regime, 
a  single scaling function could be used to describe the results:
\begin{equation}
\label{i1empscale}
|I_{1}| \sim f(L/\xi)/ L.
\end{equation}
This  behavior is similar to that of  the amplitude of the
noninteracting current given by (\ref{i0amp}).  Upon increasing the impurity
scattering further, 
the first order current  decreases
with disorder.   When the single particle wavefunctions  are sufficiently 
localized, $I_1$ can be described by a different scaling form:
\begin{equation}
\label{strongscale}
|I_{1}| \sim g(L/\xi),
\end{equation}
where  $g$  is a decreasing function of its argument.  
Eq. (\ref{strongscale}) suggests that in the localized regime $I_1$ dominates $I_0$ for large system sizes. However for localized doubly occupied single electron states 
\begin{equation}
\label{intme}
\langle \psi_{\alpha\uparrow}\psi_{\alpha\downarrow} | U \sum_i n_{i\uparrow}n_{i\downarrow}|\ \psi_{\alpha\uparrow}\psi_{\alpha\downarrow}      \rangle \sim {U\over \xi}
\end{equation}
Thus even for weak interactions, the interaction corrections may be  larger than the non interacting level spacings which go as  $1/L$. {\em This invalidates
perturbation theory in $U$ in the localized regime.}

\section{Avoided Level Crossings Theory}
\label{SecALC}
Here we will discuss how the numerical results of the previous section
can be understood  in terms of avoided level crossings (ALC)
at weak disorder.  First, we derive the ALC approximation for the noninteracting current $I_0$.

\subsection{ALC theory for $I_0$}
\label{sec:nonint}
In Fig.~\ref{spectrum} we can see a typical spectrum for a tight-binding Hamiltonian of a 6 site ring, as a  
function of the
applied flux. In the absence of disorder ($W=0$) the eigenenergies are
\begin{equation}
\label{elevels}
\epsilon_{n}(\Phi)= -2 \cos \left( \frac {2 \pi} {N} 
(n + \frac {\Phi} {\Phi_{0} } ) \right),
\end{equation}
with period $\Phi_{0}=2 \pi$.  At points of time reversal symmetry, i.e. where $\Phi$ is an integer multiple of
$\Phi_0/2$, level crossings occurs between states of opposite angular momenta. The noninteracting persistent current is a sum over the currents carried by all occupied levels,
\begin{equation}
\label{i0sum}
I_{0}(\Phi)
= - \frac { \partial E_{0} } {\partial \Phi}
= -\sum_{n,s}^{occupied} 
\frac {\partial} {\partial \Phi}\epsilon_{n}(\Phi),
\end {equation}
where $n$,$s$ are the orbital and spin index, respectively. 
The noninteracting current $I_{0}(\Phi)$ will be a smooth function of $\Phi$
away from  the points of level crossings.
By symmetry of $\Phi \to -\Phi$,  any pair of levels 
cross  with opposite slopes, and thus if they are both fully occupied their 
contribution  to the total current  cancels.  The only
nonvanishing contribution comes from a topmost level which is not compensated by its partner.  Then the  current
changes sign abrubptly as the occupation moves  from one branch 
to another.  
Thus for an odd number of fully occupied orbitals 
the current discontinuity occur
at  $\Phi~=(m + 1/2)\Phi_{0}$, otherwise the 
discontinuities will occur at $\Phi=m\Phi_0$.  In between the
discontinuities the current varies linearly with the flux, 
which explains the periodic sawtooth shape  for $I_{0}$ as seen in  
Fig.~\ref{i0plot}. 
Introducing  a small amount of disorder lifts the degeneracy at the
crossings by  opening  small 
gaps. 
This reduces the persistent current $I_0$ since the occupied levels have 
smaller slopes near the former crossing points. 
This  explains the behavior shown
in Fig.~\ref{i0plot}: weak impurity scattering softens the discontinuities
and leads to an overall 
reduction of the magnitude of current. We can quantify this observation by examining pairs of levels with momenta $\pm k$ which cross at $\Phi=0$. We specialize to the case of an even number of filled levels with equal occupations for both spin directions. The unperturbed energies are given by
\begin{equation}
\epsilon_{\pm k}(\Phi)=
 -2 \cos \left(  \pm k + \frac{ 2 \pi} {L} \frac{\Phi}{\Phi_0}\right )
\end{equation}
with $k=2 \pi  m/ L $, for a positive integer $m$. 
Consider the effect of a weak random potential $\epsilon_i$ which lifts the degeneracy in the $2\times 2$ subspace of $k,-k$,
\begin{equation}
\label{mat2X2}
H_0 \Longrightarrow
\left[
\begin{array}{cc}
	\epsilon_{k}(\Phi)  &    \tilde{V}_{k,-k}  \\
	\tilde{V}_{-k,k}   &  \epsilon_{-k}(\Phi) 
	
\end{array}
\right],
\end{equation}
where 
\begin{equation}
\label{fourier}
\tilde{V}_{k,-k}= \frac{1}{L} \sum_{i=1}^{L}   \exp(-i2k x_i) \epsilon_i.
\end{equation}
From now on we
will omit the subcripts $k,-k$ off $\tilde{V}$.

The eigenvalues and normalized eigenfunctions of (\ref{mat2X2}) are:
\begin{eqnarray}
\epsilon_{+}(\Phi):&~~~ \psi_{+}= 
\frac{1}{\sqrt{L}}\left( 
A_{+}e^{ik x}+
B_{+}e^{-ik x}\right)\nonumber\\
\epsilon_{-}(\Phi):&~~~\psi_{-}= 
\frac{1}{\sqrt{L}}\left( 
A_{-}e^{ik x}+
B_{-}e^{-ik x}\right) 
\label{ABdef}
\end{eqnarray}
where:
\begin{eqnarray} 
A_{-}&=&  e^{i \delta_{-} }    \left[ 
1 + \frac { 4|\tilde{V}|^{2} } { \left( \epsilon_{k}- \epsilon_{-k} +
\sqrt{ ( \epsilon_{-k}- \epsilon_{k})^{2} + 4|\tilde{V}|^{2}   } \;
  \right) ^{2} }   \right] ^ { -1/2}, 
\nonumber \\
B_{-} &=& \frac { -2 \tilde{V}^{*} } { 
\epsilon_{-k}- \epsilon_{k} +
\sqrt{ (\epsilon_{-k}- \epsilon_{k})^{2} + 4|\tilde{V}|^{2}   } } \; 
A_{-},
\end{eqnarray}
and,
\begin{equation}
\epsilon_{\pm} = \frac {  \epsilon_{k} + \epsilon_{-k} \pm 
\sqrt{ (\epsilon_{k} - \epsilon_{-k})^{2} + 4|\tilde{V}|^{2} }  } {2} ,
\end{equation}
which satisfy:
\begin{equation}
\epsilon_{+}(\Phi)+\epsilon_{-}(\Phi)=\epsilon_{k}(\Phi)+\epsilon_{-k}(\Phi).
\end{equation}
We see that the contribution of the  occupied orbitals to $I_0$ is
unchanged by weak disorder since:
\begin{equation}
\label{sumiab}
I_0^{k,-k}=  - \frac { \partial ( \epsilon_{+} + \epsilon_{-}) }
		{ \partial \Phi}
	=  - \frac { \partial ( \epsilon_{k} + \epsilon_{-k}) }
		{ \partial \Phi}
\end{equation}

Let us now use (\ref{i0sum}) to calculate $I_0(\Phi)$ for free
electrons with weak disorder by summing over occupied levels.  
We fill all levels up to $k_f= 2 \pi m_f/L$, 
plus two electrons at $k_f$ so that an 
even number of orbital levels is filled. 
The total noninteracting current is given by the contribution from the
filled level pairs, together with the contribution of lowest $m=0$
(nondegenerate)  level and the topmost orbital level at $m_{f}$:
\begin{eqnarray}
I_0(\Phi)&=& \sum_{m} I_{0}^{m} + I_{0}^{m=0} + 
	I_{0}^{m_f}  \nonumber \\
&=& -2\sum_{-(m_f-1)}^{m_f-1} 
	\frac {4 \pi} {L \Phi_0} \sin \frac{2\pi}{L}
	\left(m+ \frac{\Phi}{\Phi_0}\right)  
	-2 \frac{\partial \epsilon_{-}} {\partial \Phi} 
\end{eqnarray}
Neglecting corrections of order $1/L^2$, we obtain the expression:
\begin{equation}
\label{i0formula}
I_{0}(\Phi)=  \frac{8 \pi \sin(k_{f}) }{L \Phi_{0} } 
	\left[
-2 \frac{\Phi}{\Phi_{0}} +
  \left( \frac{\Phi} {\Phi_{0} } \right)   \left/ 
      \sqrt{  \left( \frac{\Phi} {\Phi_{0}} \right)^{2} +
	\left( \frac { L |\tilde{V}| } {4 \pi \sin(k_{f}) } \right) ^{2} }\right. 
\right],
\end{equation}
for $\Phi \in [-\Phi_{0}/2, \Phi_{0}/2]$.  This expression for $I_{0}$ is
valid as long as the energy scale of the disorder is much smaller than the 
level spacing at the Fermi level: 
\begin{equation}
|\tilde{V}| << 4 \pi sin(k_{f}) / L.
\label{vtilde}
\end{equation} 
Using the relation between $|\tilde{V}|$ and the mean free path $l_{el}$
defined by 
the  one dimensional Born approximation,
\begin{equation}
l_{el}~\equiv {2 \pi \sin^2 k_f\over |\tilde{V}|^2 L}  .
\label{lel-def}
\end{equation}
According to (\ref{vtilde}), the ALC approximation is  valid  for
\begin{equation}
 8 \pi l_{el}/L \ge 1,
\end{equation}
which is the ballistic, or delocalized regime.\footnote{There is no intermediate ``diffusive regime'' between the localized and ballistic regimes in one dimension\cite{loc1d}.}
In terms of $l_{el}$ one can write 
\begin{equation}
\label{i0exp}
I_{0}(\Phi)=  \frac{8 \pi \sin(k_{f}) }{L \Phi_{0} } 
	\left[
-2 \frac{\Phi}{\Phi_{0}} +
  \left( \frac{\Phi} {\Phi_{0}} \right)   \left/ 
      \sqrt{\left(\frac{\Phi} {\Phi_{0}}\right)^{2} +
	 \frac { L } {8 \pi l_{el} }  }   \right.
\right].
\end{equation}
In order to compare (\ref{i0exp}) to the numerical result for Eq. (\ref{h0}),
one needs to determine the parameter $l_{el}$. Since fluctuations in $I_0$ for different disorder realizations, we determine
$l_{el}$ by fitting (\ref{i0exp}) to the numerical $I_0$. Once $l_{el}$
is determined for a particular disorder realization, we use it to evaluate $I_1$ as shown below.

\subsection{ALC theory for  $I_1$}
We shall now proceed to use the same approximation to explain the
behavior of $I_1$.\footnote{This approach is similar  to the discussion of
Aharonov-Bohm  oscillations of the participation ratio in disordered rings Ref.~\cite{nagel}.}  

According to (\ref{e1def}), the first order energy is a sum of the density
squared at all sites, 
\begin{equation}
E_{1}(\Phi)= U \sum_{i} n_{i \uparrow} n_{i \downarrow}
\end{equation}
where for the filling of an even number of orbitals per spin, the single spin  density is,
\begin{equation}
\label{densdef}
n_{is}(\Phi)= 
 |\psi_{0} (i)|^{2}+|\psi_{m_f,-} (i)|^{2}~+\sum_{m=1}^{ m_f-1} \sum_{\sigma=\pm} |\psi_{m,\sigma} (i)|^{2}.
\end{equation}
We will now show that the first order
energy,
\begin{equation}
E_{1}(\Phi)= U \sum_{i} n_{i \uparrow} n_{i \downarrow}
\end{equation}
is enhanced near level 
crossings. 
 
For a pair of fully occupied levels one has, by unitarity
\begin{equation}
|\psi_{m,+}(i)|^2 + |\psi_{m,-}(i)|^2 = |\psi_{k}(i)|^2 + |\psi_{-k}(i)|^2={2\over L}.
\end{equation}
Consequently, for weak disorder, all occupied levels, 
except for the last one, contribute constant values to
the total density: 
\begin{equation}
\label{rhoiup}
n_{i\uparrow}= \frac{2m_f}{L} + |\psi_{-}(i)|^2
\end{equation}
or in terms of the coefficients $A,B$ (\ref{ABdef})
\begin{equation}
\label{dens}
n_{i\uparrow}=\frac{2m_f+1}{L} + 
\frac{ A_{-}^{*} B_{-} e^{i2k x_i} } { L } + 
\frac{ A_{-} B_{-}^{*} e^{-i2k x_i} } { L }. 
\end{equation}
The first order  energy is thus given by (\ref{rhoiup}),
\begin{equation}
\label{densfluc}
E_{1}(\Phi)/U 
= \frac{ (2m_f+1)^{2} } {L}
+ \frac{ 2 |A_{-}|^{2} |B_{-}|^{2} } {L} .
\end{equation}
Using (\ref{ABdef}) and (\ref{lel-def}) we can write 
\begin{equation}
\label{e1phi}
E_{1}(\Phi)/U 
\approx \frac{ (2m_f+1)^{2} } {L} + 
\frac{1} {2 L} 
\left[ 1 +  \left( \frac{8  \pi l_{el})  }{ L } \right) 
	\left(\frac{\Phi}{\Phi_{0}} \right)^{2}  \right] ^ {-1} ,
\end{equation}
Differentiating (\ref{e1phi}) with respect to flux yields
\begin{equation}
I_{1}(\Phi)/U ~=
{8\pi l_{el} \Phi\over (L\Phi_0)^2}~\left/ 
\left[ 1 +  \left( \frac{8 \pi l_{el} } {L} \right) 
	\left(\frac{\Phi}{\Phi_{0}} \right) ^{2}  \right]^ {2}\right. .
\label{i1anal}
\end{equation}
In Fig.~\ref{i1theory}, (\ref{i1anal}) is compared to the numerical result
for $I_1$ for various
values of disorder. For each disorder realization, $l_{el}$ is determined by fitting the numerical and ALC results for $I_0$.  We see that for
weak disorder there is a satisfying agreement between the ALC approximation and the numerical
results. In Fig. \ref{i1theory}c,   the disorder is too
large, and the ALC approximation fairs badly. 

Eq. (\ref{i1anal}) explains  both the positive correlation between $I_0$ and $I_1$  of 
(\ref{poscor}).  Since in one dimension
the mean free path ($l_{el}$) and the localization length ($\xi$)
differ by a proportionality factor of order unity \cite{loc1d},  (\ref{i1anal})  agrees with the empirical scaling form  (\ref{i1empscale}) $|I_{1}| \sim f(L/l_{el})/ L$ which was found numerically at weak disorder.

\section{Difference of Hubbard Model and Spinless Fermions}
\label{Sec4}
In recent papers \cite{berkovits,nnexact}, small disordered rings of spinless electrons have been exactly diagonalized. The persistent currents have been computed as a function of disorder and interaction strength. In contrast to our  
result for the Hubbard model, interactions have were seen to {\it reduce}
the persistent current at weak disorder.  Here we apply the ALC
approach to explain this apparent difference between the models.

The spinless Hamiltonian $H^{s}$ is given by 
\begin{eqnarray}
 H^{s}&=& H^{s}_{0} \; + \;    \sum_{i} U_{ij} n_{i} n_{j},\nonumber\\
 H^{s}_{0}&=& - \sum_{i}[  e^{i( \frac{ 2 \pi} {L} \frac{\Phi} {\Phi_0} )} 
a_{i+1}^{\dagger}  a_{i}+{\rm h.c.}]
\; + \; \sum_{i } \epsilon_{i} a_{i}^{\dagger} a_{i}. 
\label{s}
\end{eqnarray}
The non interacting current, $I^{s}_0$, is given by half the value of Eq. (\ref{i0sum}). In the ALC approximation,  it is given by half the value of
Eq. (\ref{i0exp}). 

Following the analogous derivation of $I_1$, we  
use (\ref{dens}) to obtain:
\begin{equation}
E^{s}_{1}(\Phi)=  (2m_f+1)^{2}  \tilde{U}(0)
+ 2 |A_{-}|^{2} |B_{-}|^{2} \tilde{U}(2 k_{f}),
\end{equation} 
where 
\begin{equation}
\tilde{U}(k)= \frac{1}{L} \sum_{j \neq i} \exp(-i k x_j) U_{ij},
\end{equation}
which implies
\begin{eqnarray}
E^{s}_{1}(\Phi)~&=& { (2m_f+1)^{2} } \tilde{U}(0)  +
\tilde{U}(2 k_{f})
\left[ 1 +  \left( \frac{8 \pi l_{el} } {L} \right) 
	\left(\frac{\Phi}{\Phi_{0}} \right) ^{2}  \right] ^ {-1}\nonumber\\
I^{s}_1(\Phi)~&=&{ 16 \pi l_{el}   \Phi
\over \Phi^2_{0} L }  \tilde{U}(2 k_{f}) \left[ 1 +  \left( \frac{8 \pi l_{el} } {L} \right) 
	\left(\frac{\Phi}{\Phi_{0}} \right) ^{2}  \right] ^ {-2} 
\label{e1general}
\end{eqnarray}

For the spinless nearest  neighbor case studied in Ref. \cite{nnexact}:
\begin{equation}
E_1(\Phi)= U \sum_{i} n_{i} n_{i+1},
\end{equation}
the corresponding Fourier coefficient is
\begin{equation}
\tilde{U}(2 k_{f})= 2 U\cos (2 k_f).  
\end{equation}
Above a quarter filling, $k_f \ge \pi/4$, ${\tilde U}$ is  negative. Consequently $I^{s}_1$ has the opposite sign to that of $I_1$ in Eq. (\ref{i1anal}), and to $I^{s}_0$. That is to say,  the persistent current of the spinless fermion model is 
{\em suppressed} by the interactions.

The difference between the models can be attributed to the different effects  of {\em intersite} interactions (in the spinless model), versus {\em local}  interactions (in the Hubbard model).

\section{Remarks and Conclusions}

We have investigated the first order effect of  Hubbard interactions on the 
persistent current in one dimensional disordered rings.

The findings can be summarized as follows.  $I_1$  was found to correlate in sign
with  $I_0$, a fact which goes contrary to the naive intuition that interactions suppress currents (e.g. as they do for the conductivity
of the Luttinger model with an impurity \cite{Kane}). 

We can understand this result by observing that $I_1$ depends on charge fluctuations which vary strongly with flux near the degeneracy points, which also determine the sign and magnitude of $I_0$. Using the avoided level  crossings
theory, we obtain analytical expressions which fit the numerical results for  both $I_0$ and $I_1$. We can use the same theory to explain why spinless fermions with intersite interactions exhibit suppression of currents rather than enhancement.
 
We have found numerically that $I_1(L/\xi)$ scales differently with the localization length 
than $I_0(L/\xi)$, it seems that effects of
electron-electron interactions grow as  disorder
is increased.

However, first order perturbation theory in the weakly disordered case is expected to hold as long as the interaction matrix elements do not exceed
the single particle level spacings. Thus we are restricted to the regime
$U\le 1$. We can draw on the exact diagonalization results  \cite{berkovits,nnexact} where for weak disorder, the numerical currents
are found to  vary linearly with interaction strength in a sizeable regime. 
Thus we believe
that first order perturbation theory should be valid for physically interesting  interaction parameters.

It would be very satisfying if a similar analysis could be applied
to  the experimentally relevant case of three dimensional  rings. 
Preliminary results yield positive correlations between $I_0$ and $I_1$, but
where a simple minded application of the ALC approach cannot describe the diffusive regime of 
$l_{el} << L$.
It would also be important to understand the effects of true long range Coulomb interactions with the screening and exchange effects which are
absent
in the Hubbard model.

\subsection*{Acknowledgements}
We thank E. Akkermans, R. Berkovits and Y. Gefen for useful discussions.
The support of the US-Israel Binational Science Foundation,
the Israeli Academy of Sciences and the Fund for Promotion of Research at Technion is gratefully acknowledged. Part of this work has been supported by grant from US Department of Energy, DE-FG02-91ER45441.  M.K. thanks the Institute for Theoretical Physics at Technion for its hospitality.

\clearpage

\begin{figure}[h]

\caption{
The noninteracting current ($I_0$) as a function of the applied flux
for  several strengths of disordered potential:
$W$=0, 0.5, 1.0, 2.0 and 4.0.  All curves are for a half-filled lattice
of six sites.
}
\label{i0plot}
\end{figure}

\begin{figure}[h]

\caption{
The first order interacting current ($I_1$) as a function of the applied flux
for  disordered potential values:
$W$= 1.0, 2.0 and 4.0. 
As before, all curves are for a half-filled lattice of six sites.
}
\label{i1plot}
\end{figure}

\begin{figure}[h]

\caption{
The logarithm of the localization length ($\ln \xi$), implied by 
Eq. (\protect\ref{i0amp}), is plotted as a function of strength of disorder
($W$) for half-filled rings  with  $N$=6, 10, 14, 20 and 25 sites.  Data
for the different system sizes collapse well onto a single curve.
}
\label{loclength}
\end{figure}

\begin{figure}[h]

\caption{
Energy level spectrum of the tight-binding Hamiltonian(\protect\ref{elevels})
as a function of applied flux for a $N$=6 ring in the absence of disorder 
(solid lines) and for weak disorder (dotted lines).
}
\label{spectrum}
\end{figure}

\begin{figure}[h]
\caption{
Comparison between numerically determined interactions correction  $I_1$
(solid lines), and the analytic ALC result (\protect\ref{i1anal}) (dashed lines). 
Figs. (a) -- (c) show results for three values of disorder strength $W$.
$l_{el}(W)$ are determined by fitting Eq. (\protect\ref{i0exp}) to the numerical disorder averaged $I_0(\Phi)$. Error bars depict fluctuations of numerical $I_1$ for different disorder realizations.
} 
\label{i1theory}
\end{figure}

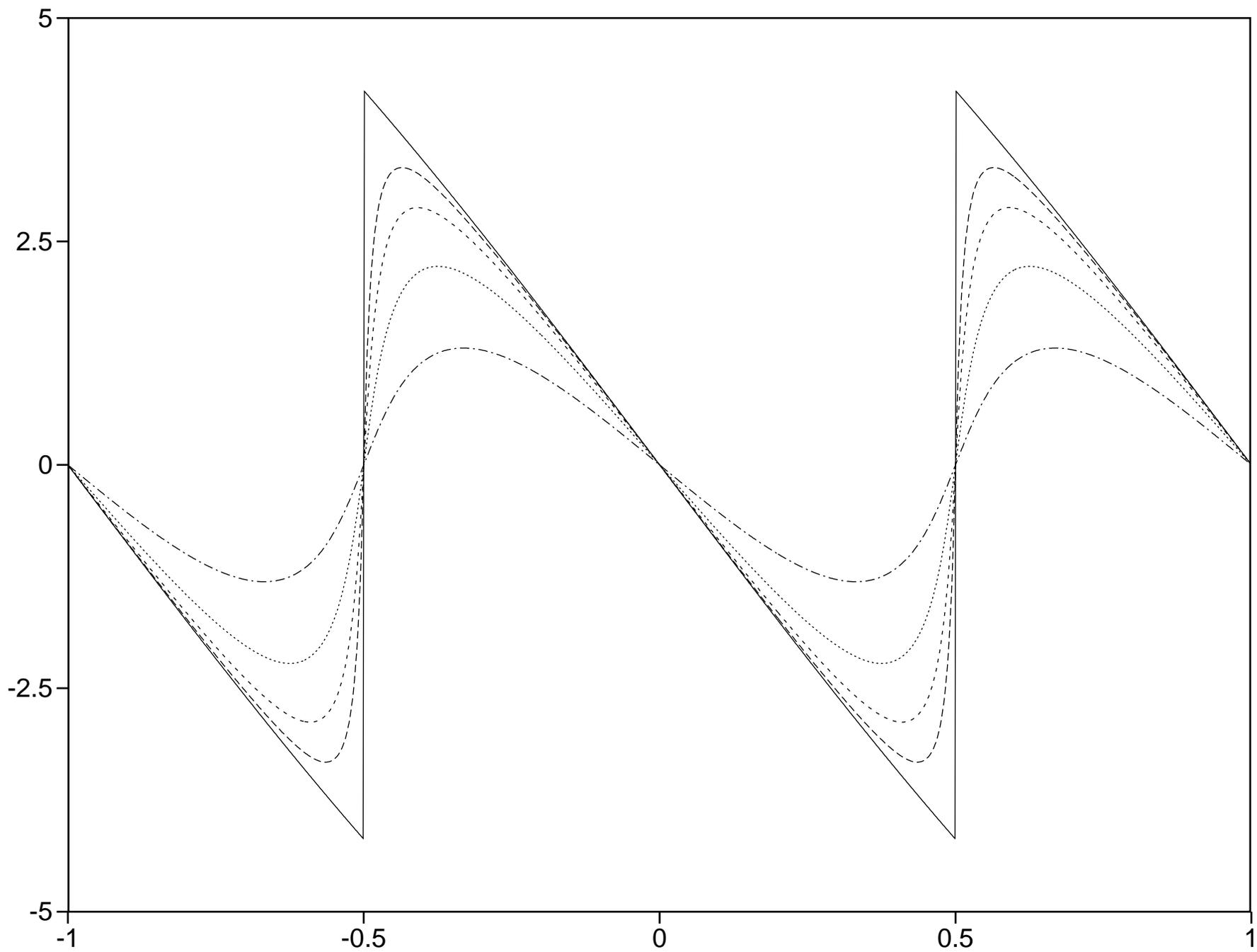
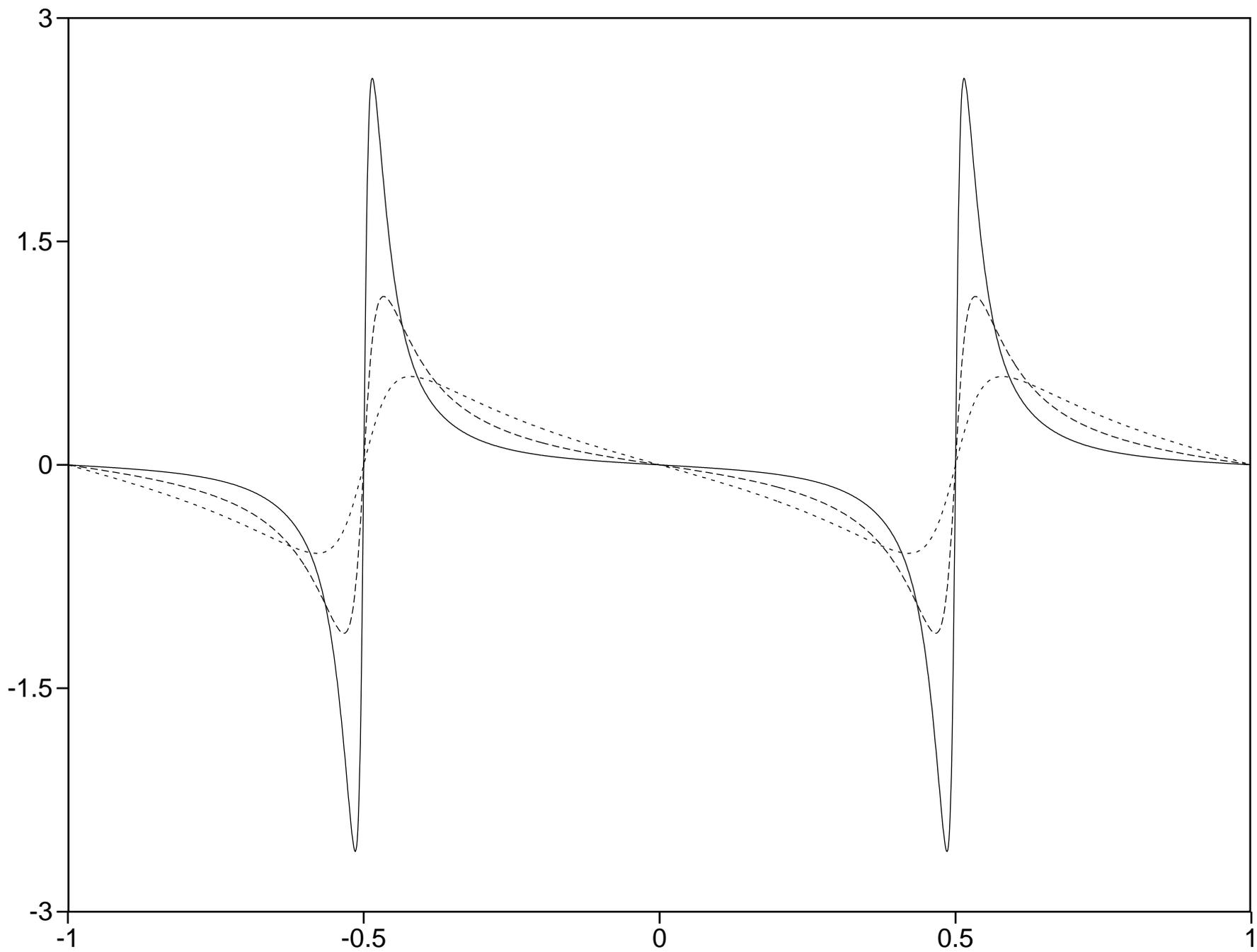
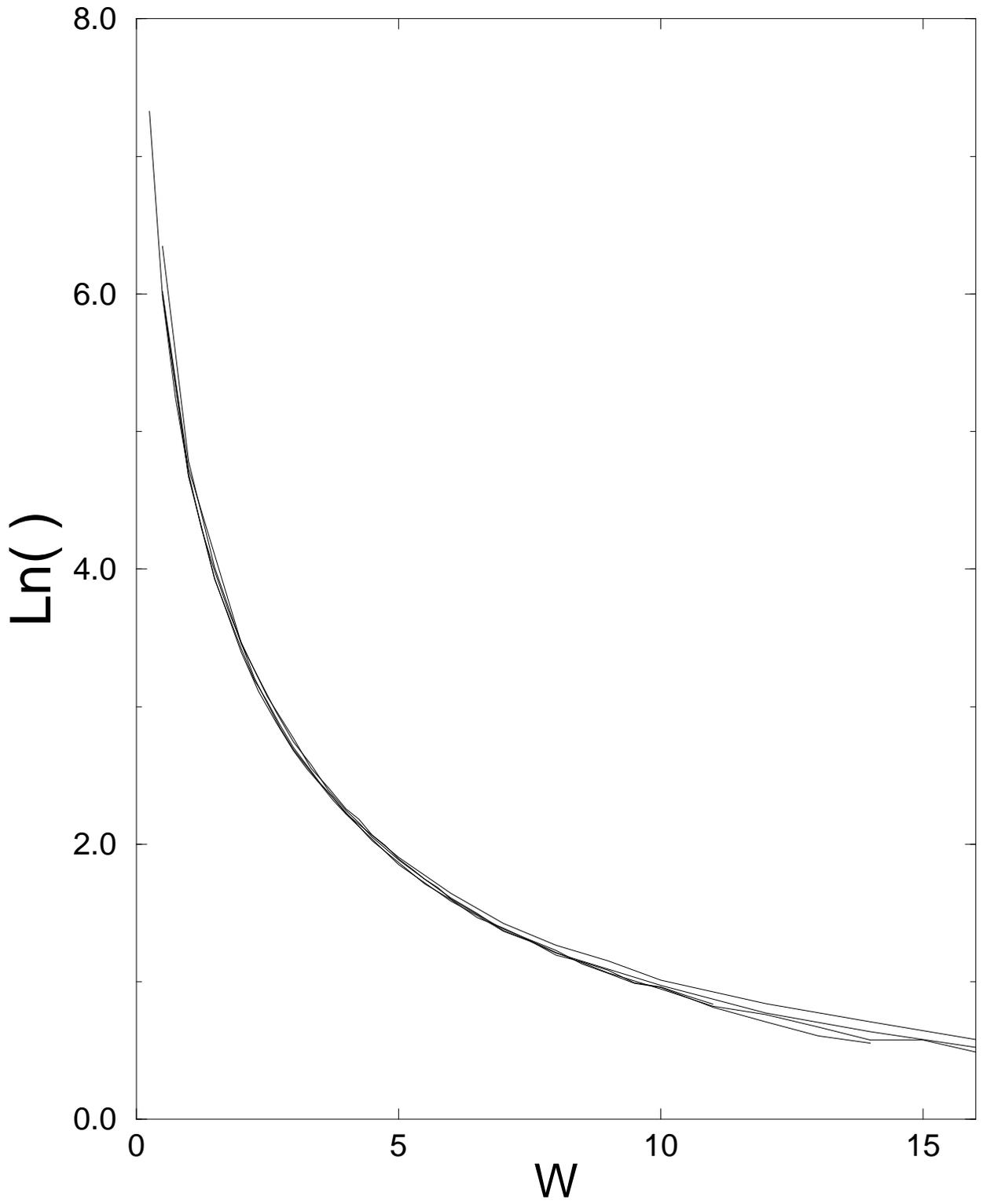
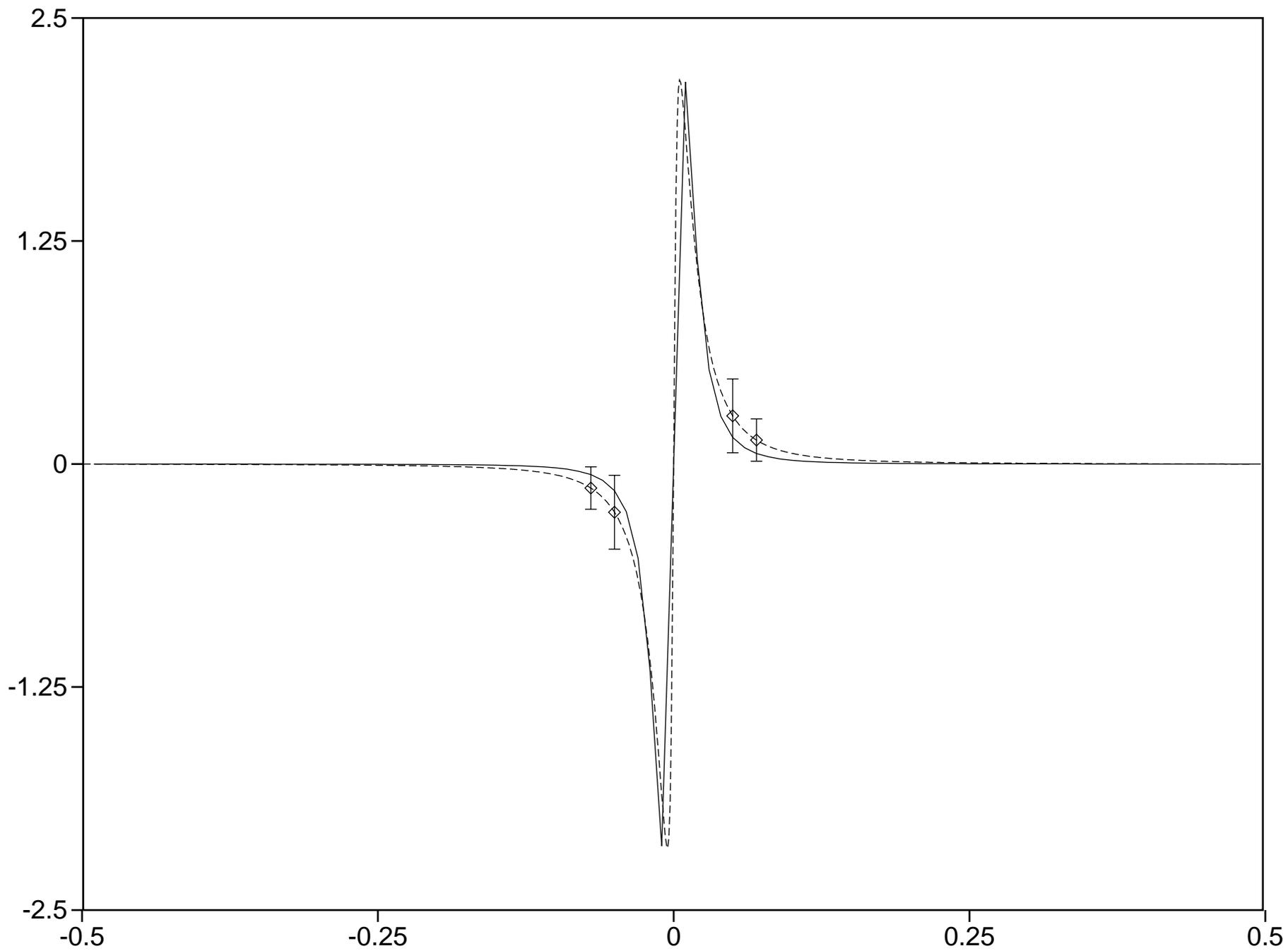
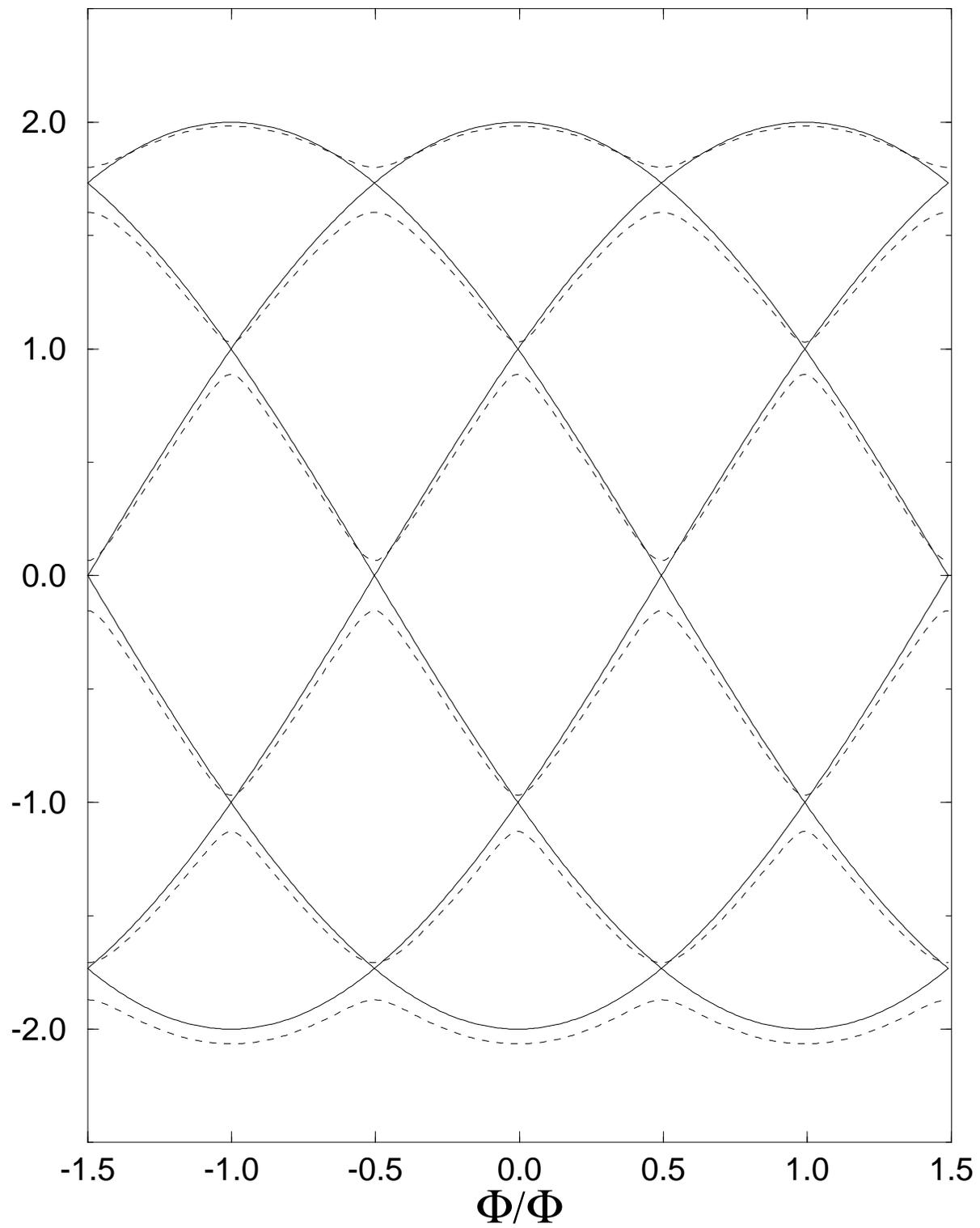
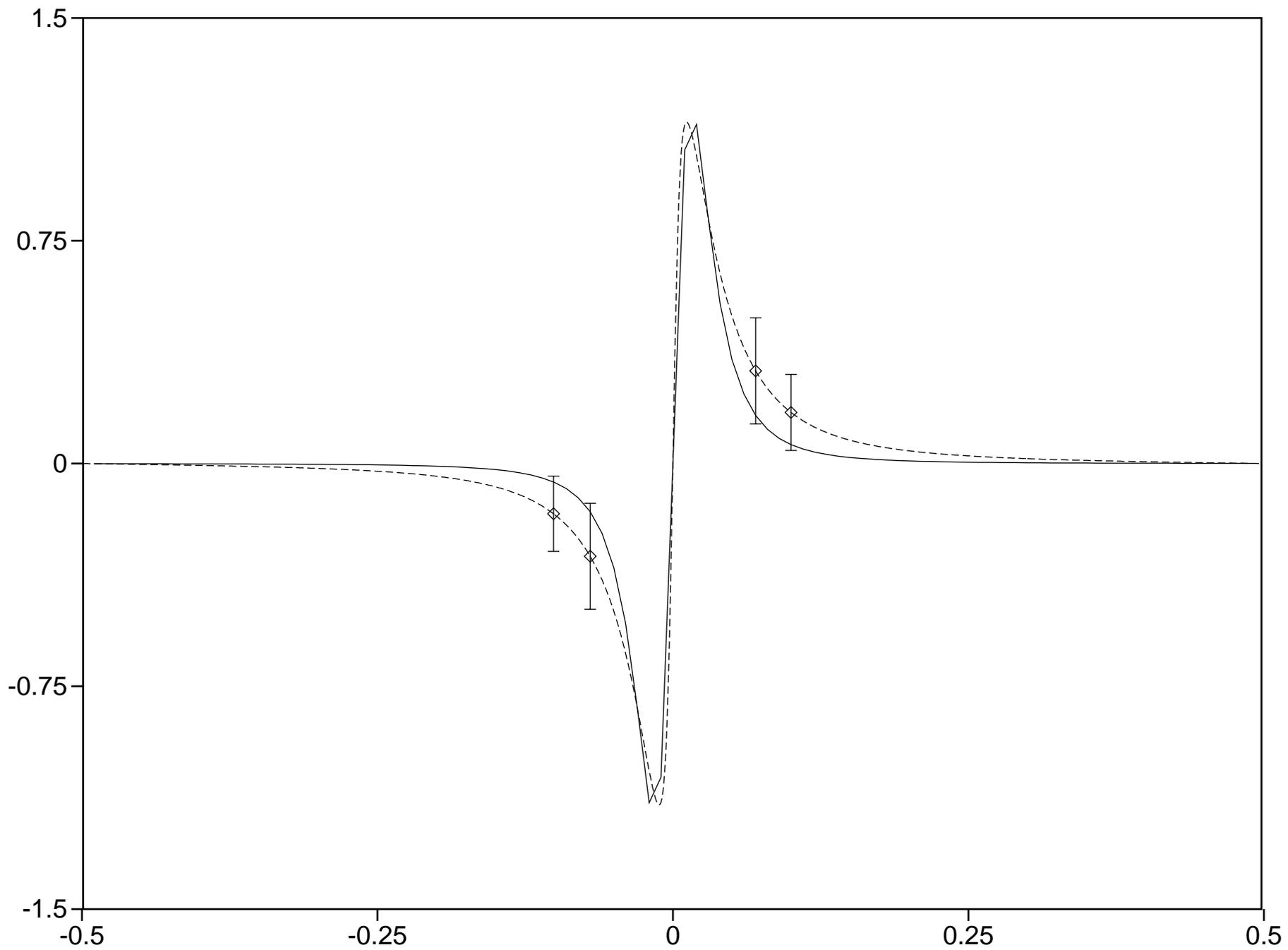
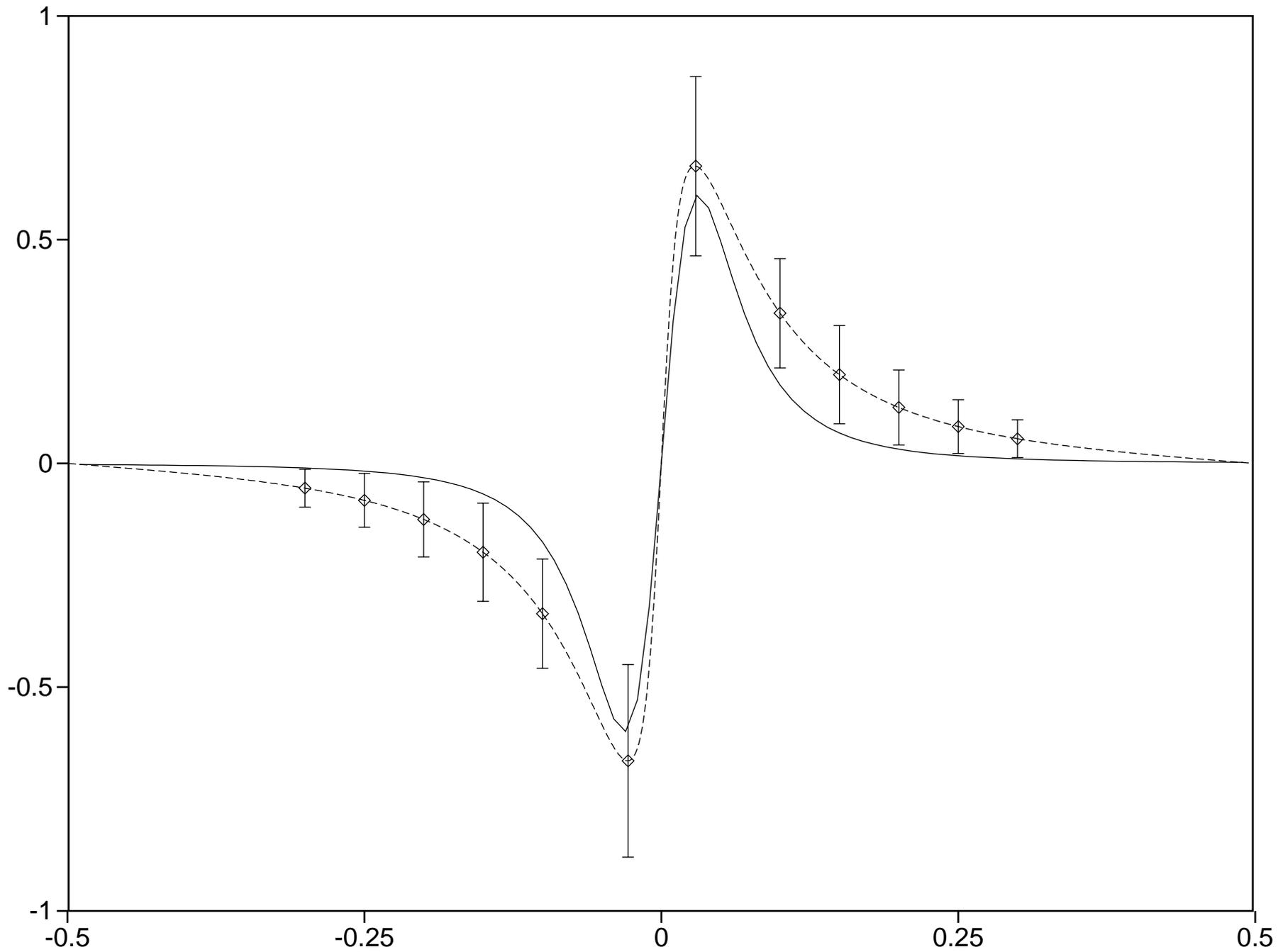
\end{document}